\documentclass[aps,prb,twocolumn,showpacs]{revtex4}
\usepackage{amsmath}
\usepackage{graphicx,epsfig,psfrag}
\usepackage{amssymb}
\usepackage{bm}

\newcommand{\be}{\begin{equation}}
\newcommand{\ee}{\end{equation}}

\newcommand{\w}{\omega}
\newcommand{\s}{\sigma}
\newcommand{\ra}{\rightarrow}
\renewcommand{\Re}{\mathrm{Re}\;}
\renewcommand{\Im}{\mathrm{Im}\;}
\renewcommand{\vec}[1]{{\bf #1}}

\begin{document}

\title{Lower bounds for the conductivities of correlated quantum systems}
\author{Peter Jung}
\author{Achim Rosch}
\affiliation{Institute for Theoretical Physics, University of
Cologne, 50937 Cologne, Germany.}
\begin{abstract}
We show how one can obtain a lower bound for the
electrical, spin or heat conductivity of correlated quantum systems
described by Hamiltonians of the form  $H=H_0+g H_1$. Here $H_0$
is an {\em interacting} Hamiltonian characterized by conservation
laws which lead to an infinite conductivity for $g=0$. The small
perturbation $g H_1$, however, renders the conductivity finite at
finite temperatures. For example, $H_0$ could be a continuum field
theory, where momentum is conserved, or an integrable
one-dimensional model while $H_1$ might describe the effects of
weak disorder. In the limit $g \to 0$, we derive lower bounds for
the relevant conductivities and show how they can be improved
systematically using the memory matrix formalism. Furthermore, we
discuss various applications and investigate under what conditions
our lower bound may become exact.
\end{abstract}
\date{\today}
\maketitle

\section{Introduction}
Transport properties of complex materials are not only important for
many applications but are also of fundamental interest as their study
can give insight into the nature of the relevant quasi particles and
their interactions.

Compared to thermodynamic quantities, the transport properties of
interacting quantum systems are notoriously difficult to calculate
even in situations where interactions are weak. The reason is that
conductivities of non-interacting systems are usually infinite even at
finite temperature, implying that even to lowest order in perturbation
theory an infinite resummation of a perturbative series is mandatory.
To lowest order this implies that one usually has to solve an integral
equation, often written in terms of (quantum-) Boltzmann equations or
-- within the Kubo formalism -- in terms of vertex equations. The
situation becomes even more difficult if the interactions are so
strong that an expansion around a non-interacting system is not
possible. Also numerically, the calculation of zero-frequency
conductivities of strongly interacting clean systems is a serious
challenge and even for one-dimensional systems reliable calculations
are available for high temperatures
only\cite{zotos.prelovsek:1996,fabricius.mccoy:1998,
  narozhny:millis.andrei:1998,zotos.prelovsek:2003,heidrich-meissner:2005-b,
  jung.rosch:2006}.

Variational estimates, e.g.~for the ground state energy, are powerful theoretical
techniques to obtain rigorous bounds on physical quantities. They
can be used to guide approximation schemes to obtain simple
analytic estimates and are sometimes the basis of sophisticated
numerical methods like the density matrix renormalization group
\cite{schollwoeck:2005}.

Taking into account both the importance of transport quantities
and the difficulties involved in their calculation it would be
very useful to have general variational bounds for transport
coefficients.

A well known example where a bound for  transport quantities has
been derived is the variational solution of the Boltzmann
equation, discussed extensively by Ziman\cite{ziman:1960}. The
linearized Boltzmann equation in the presence of a static electric
field can be written in the form 
\be 
e \vec{E} \vec{v}_{\vec{k}}\frac{d f^0}{d \epsilon_\vec{k}}
=\sum_{\vec{k}'} W_{\vec{k},\vec{k}'} \Phi_{\vec{k}'}
\ee
where $W_{\vec{k},\vec{k}'}$ is the integral kernel describing the
scattering of quasiparticles and we have linearized the Boltzmann
equation around the Fermi (or Bose) distribution
$f^0_\vec{k}=f^0(\epsilon_\vec{k})$ using
$f_{\vec{k}}=f^0_\vec{k}-\frac{d f^0}{d \epsilon_\vec{k}}
\Phi_\vec{k}$.  Therefore, the current is given by $\vec I=-e \sum_\vec{k}
\vec{v}_{\vec{k}} \frac{d f^0}{d \epsilon_\vec{k}} \Phi_\vec{k}$
and the dc conductivity is determined from the inverse of the
scattering matrix $W$ using $\sigma=-e^2 \sum_{\vec{k}\vec{k}'}
\frac{d f^0}{d \epsilon_\vec{k}} {v}^i_{\vec{k}}
W^{-1}_{\vec{k},\vec{k}'} {v}^i_{\vec{k}'} \frac{d f^0}{d
\epsilon_{\vec{k}'}}$. It is easy to see that this result can be
obtained by maximizing a 
functional\cite{kohler:1948,kohler:1949,sondheimer:1950,ziman:1960}
$F[\Phi]$ with
\begin{eqnarray}
\sigma&=&e^2 \max_{\Phi} F[\Phi]\ge e^2 \max_{a_i} 
F\left[\sum_i a_i \phi_i\right] \label{varBoltzmann}\\
F[\Phi]&=& \frac{2 \left( \sum_\vec{k} {v}^i_{\vec{k}} \Phi_\vec{k}
\frac{d f^0}{d \epsilon_\vec{k}}\right)^2}{\sum_{\vec k,\vec k'}
(\Phi_\vec{k}-\Phi_{\vec{k}'})^2 W_{\vec{k},\vec{k}'}} \nonumber
\end{eqnarray}
where we used that $\sum_{\vec k'}
W_{\vec{k},\vec{k}'}=0$ reflecting the conservation of
probability. The variational formula (\ref{varBoltzmann}) is
actually closely related\cite{ziman:1960} to the famous H-theorem of
Boltzmann which states that entropy always increases upon
scattering.

A lower bound for the conductivity can be obtained by varying
$\Phi$ only in a subspace of all possible functions. This
allows for example to obtain analytically good estimates for
conductivities {\em without} inverting an infinite dimensional
matrix or, euqivalently, solving an integral equation, see Ziman's
book for a large number of examples\cite{ziman:1960}.

The applicability of Eq.~(\ref{varBoltzmann}) is restricted to
situations where the Boltzmann equation is valid and bounds for
the conductivity in more general setups are not known. However,
for ballistic systems with infinite conductivity it is possible to
get a lower bound for the so-called Drude weight. Mazur
\cite{mazur:1969} and later Suzuki \cite{suzuki:1971} considered situations
where the presence of conservation laws prohibits the decay of
certain correlation functions in the long time limit. In the
context of transport theory their result can be applied to systems
(see Appendix~\ref{appA} for details) where the finite-temperature
conductivity $\sigma(\omega,T)$ is {\em infinite}  for $\omega=0$
and characterized by a finite Drude weight $D(T)>0$ with
\begin{equation}
\Re\sigma(\omega,T)=\pi D(T) \delta(\omega)+\sigma_{\rm
reg}(\omega,T). \label{drude}
\end{equation}
Such a Drude weight can arise only
in the presence of exact conservation laws $C_j$ with $[H,C_j]=0$.
Suzuki \cite{suzuki:1971} showed that the Drude weight can be expressed
as a sum over {\em all} $C_j$
\begin{equation}
D=\frac{\beta}{V}\sum_{j=0}^\infty  \frac{\langle C_j J
\rangle^2}{\langle C_j^2 \rangle}\geq\frac{\beta}{V}\sum_{j=0}^N
\frac{\langle C_j J \rangle^2}{\langle C_j^2
\rangle}.\label{DrudeMazur}
\end{equation}
where $J$ is the current associated with $\sigma$. 
For convenience a basis in the space of $C_i$ has been
chosen such that $\langle C_i C_j \rangle=0$ for $i \neq j$. More
useful than the equality in Eq.~(\ref{DrudeMazur}) is often the
inequality\cite{mazur:1969} which is obtained when the sum is
restricted to a finite subset of conservation laws. Such a finite
sum over simple expectation values can often be calculated rather
easily using either analytical or numerical methods. The Mazur
inequality has recently been used
heavily \cite{zotos:1997,fujimoto.kawakami:2003, %
zotos.prelovsek:2003,heidrich-meissner:2005,sakai:2005} to
discuss the transport properties of one-dimensional
systems.

Model systems, due to their simplicity, often exhibit symmetries not
shared by real materials. For example, the heat conductivity of
idealized one-dimensional Heisenberg chains is infinite at
arbitrary temperature as the heat current is conserved. However,
any additional coupling (next-nearest neighbor, inter-chain,
disorder, phonon,...) renders the conductivity
finite\cite{zotos.prelovsek:1996, zotos.prelovsek:2003, jung.rosch:2006,shimshoni:2003,heidrich-meissner:2005-b,heidrich-meissner:2002,rozhkov.chernyshev:2005}. 
If these perturbations are weak, the heat
conductivity is, however, large as observed
experimentally\cite{sologubenko:2000,hess:2007}. For a more general
example, consider
an arbitrary translationally invariant
continuum field theory. Here  momentum is conserved which usually
implies that the conductivity is infinite for this model. In real
materials momentum decays by Umklapp scattering or disorder
rendering the conductivity finite. It is obviously desirable to
have a reliable method to  calculate transport in such situations.
\begin{figure} \begin{center} 
\includegraphics[width=.98\linewidth,clip=]{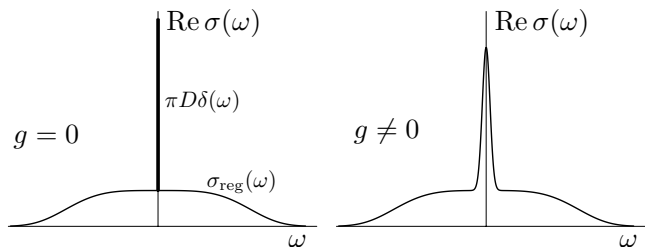} 
\end{center} \caption{\label{drudepic}For $g=0$ a Drude peak shows up in the conductivity, resulting from exact conservation laws. For $g\neq 0$ the Drude peak broadens and the dc conductivity becomes finite.}
\end{figure}
In this work we consider systems with the Hamiltonian \be H=H_0+g H_1,
\label{h} \ee where for $g=0$ the relevant heat-, charge- or spin-
conductivity is infinite and characterized by a finite Drude weight
given by Eq.~(\ref{DrudeMazur}). As discussed above, $H_0$ might be an
integrable one-dimensional model, a continuum field theory, or just a
non-interacting system. The term $g H_1$ describes a (weak)
perturbation which renders the conductivity finite, e.g. due to
Umklapp scattering or disorder, see Fig.~\ref{drudepic}. Our goal is
to find a variational lower bound for conductivities in the spirit of
Eq.~(\ref{varBoltzmann}) for this very general situation, without any
requirement on the existence of quasi particles. For technical reasons
(see below) we restrict our analysis to situations where $H$ is time
reversal invariant.

In the following, we first describe the general setup and
introduce the memory matrix formalism, which allows us to formulate
an inequality for transport coefficients for weakly perturbed
systems. We will argue that the inequality is valid under the
conditions which we specify. Finally, we investigate under which
conditions the lower bounds become exact and briefly discuss
applications of our formula.

\section{Setup}\label{setup}

Consider the local density $\rho(x)$ of an  \emph{arbitrary}
physical quantity which is locally conserved, thus obeying a
continuity equation $$\partial_t \rho+\nabla j=0.$$ Transport of
that quantity is described by the dc conductivity $\s$ which
is the response of the current to some external field $E$ coupling
to the current,
\[\langle J \rangle=\s E,\]
where $J=\int j(x)$ is the total current and  $\langle J\rangle$
its expectation value. Note that $J$ can be an electrical-, spin-,
or heat current and $E$ the corresponding conjugate field
depending on the context. The dynamic conductivity $\s(z)$ is
given by Kubo's formula, see Eq.~(\ref{kubo}). We are 
interested in the  dc conductivity $\s=\lim_{\w\ra 0}\s(z=\w+i0)$.

Starting from the Hamiltonian (\ref{h}) we  consider a system
where $H_0$ posesses a set of exact conservation
laws $\{C_i\}$ of which at least one correlates with the current,
$\langle J C_1\rangle\neq0$. Without loss of generality we assume
$\langle C_i C_j \rangle=0$ for $i \neq j$. For $g=0$ the
Drude weight $D$ defined by Eq.~(\ref{drude}) is given by
Eq.~(\ref{DrudeMazur}). We can split up the current under
consideration into a part which is parallel to the $C_i$ and one
that is orthogonal,
$$J=J_\parallel+J_\perp,$$
with $J_\|=\sum_i \frac{\langle C_i J\rangle}{\langle
C_i^2\rangle} C_i$, which results in a separation of the conductivity,
\be
\s(z)=\s_\parallel(z)+\s_\perp(z).\label{condcomp}
\ee
Since the conductivity $\s(z)$ is given by a current-current correlation function and 
the current $J_\|$ ($J_\perp$) is diagonal (off-diagonal) in energy, 
cross-correlation functions $\langle\langle
J_\|;J_\perp\rangle\rangle$ vanish in Eq.~(\ref{condcomp}).

According to Eq.~(\ref{DrudeMazur}), the Drude peak of the unperturbed
system, $g=0$, arises solely from $J_\parallel$:
\be \Re\s_\parallel(\w)=\pi D \delta(\w),\ee 
while $\s_\perp(z)$ appears in Eq.~(\ref{drude}) as the
regular part, $\Re \s_\perp(\w)=\sigma_{\rm reg}(\w)$.

In this work we will focus on $\s_{\|}(\w)$, 
since the small perturbation is not going to affect $\s_\perp(\w)$ much
(which is assumed to be free of singularities here, see section \ref{central}) 
while $\sigma_\|(\w=0)$ diverges for $g \to 0$, see Fig.~\ref{drudepic}. 
As we are interested in the small $g$ asymptotics only,
we may neglect the contribution $\s_\perp(0)$ to the dc conductivity.
Hence we set $J=J_\parallel$ and
$\sigma(\w)=\sigma_\|(\w)$ in the following.

\section{Memory matrix formalism}

We have seen that  certain conservation laws of $H_0$ play  a
crucial role in determining the conductivity of both the
unperturbed and the perturbed system. In the presence of a small
perturbation $g H_1$, these modes are not conserved anymore but at
least some of them decay slowly. Typically, the conductivity of
the perturbed system will be determined by the dynamics of these
slow modes. To separate the dynamics of the slow modes from the
rest, it is convenient to use a hydrodynamic approach based on the
projection of the dynamics onto these slow modes. In this section we
will therefore review the so called memory matrix formalism
\cite{forster:1975}, introduced by Mori and
Zwanzig\cite{mori:1965,zwanzig:1960} for this purpose. 
In the next section we will show that
this approach can be used to obtain a lower bound for the dc
conductivity for small $g$.

We start by defining a scalar product in the space of
quantum mechanical operators,
\be
(A|B)=\int_0^\beta d\lambda \langle A^\dag
B(i\lambda)\rangle-\beta \langle A^\dag\rangle\langle
B\rangle\label{scalarProduct}
\ee
 As the next step we choose a -- for the moment
-- arbitrary set of operators $\{C_i\}$. In most applications, the
$C_i$ are the relevant slow modes of the system. For notational
convenience, we assume that the $\{C_i\}$ are orthonormalized, 
\be
(C_i|C_j)=\delta_{ij}.\label{normC}
\ee 
In terms of these we may
define the projector $P$ onto (and $Q$ away from) the space
spanned by these `slow' modes
$$P=\sum_i |C_i)(C_i|=1-Q.$$
We assume that $C_1$ is the current we are interested in,
$$|J)\equiv|C_1).$$
The time evolution is given by the Liouville-(super)operator
$$L=[H,.]=L_0+g L_1$$
with $(LA|B)=(A|LB)=(A|L|B),$ and the time
evolution of an operator may be expressed as $|A(t))=|e^{i H
t}Ae^{-i H t})=e^{i L t}|A).$ With these notions, one obtains the
following simple, yet formal expression for the conductivity:
$$\s(\w)=\left(J\left|\frac{i}{\w-L}\right|J\right)= \left(C_1\left|\frac{i}{\w-L}\right|C_1\right).$$
Using a number of simple manipulations, one can
show\cite{mori:1965,zwanzig:1960,forster:1975} that  the conductivity can be
expressed as the $(1,1)$-component of the inverse of a matrix
\begin{equation}
\s(\w)= \left( {M}(\w)+i {K}-i \w
\right)^{-1}_{\phantom{-1}11},\label{condty}
\end{equation} 
where
\be
{M}_{ij}(\w)=\left(\dot{C}_i{\Big|}Q\frac{i}{\w-LQ}{\Big|}\dot{C}_j\right)\label{memdef}
\ee
is the so-called memory matrix and
\be
{K}_{ij}=\left(\dot{C}_i|{C}_j\right)\label{Kdef}
\ee
a frequency independent matrix. The formal expression
(\ref{condty}) for the conductivity is exact, and completely
general, i.e.~valid for an arbitrary choice of the modes $C_i$
(they do not even have to be `slow'). Only $C_1=J$ is required.
However, due to the projection operators $Q$, the memory matrix
(\ref{memdef}) is in general difficult to evaluate. It is when one
uses approximations to ${M}$ that the choice of the projectors
becomes crucial (see below).

Obviously, the dc conductivity is given by the $(1,1)$-component
of 
\be
({M}(0)+{K})^{-1}.\label{olimit}
\ee
More generally, the $(m,n)$-component of Eq.~(\ref{olimit}) describes
the response of the `current' $C_m$ to an external field coupling
solely to $C_n$. We note, that since a matrix of transport
coefficients has to be positive (semi)definite, this also  holds
for the matrix ${M}(0)+{K}$.

To avoid technical complications associated with the presence of
${K}$ we restrict our analysis in the following to
time reversal invariant systems and choose the $C_i$ such, that
they have either signature $+1$ or $-1$ under time reversal 
\footnote{As $\Theta^2=\pm 1$ for states with integer or half-integer spin, 
the combinations $A\pm \Theta A\Theta^{-1}$ have signatures $\pm 1$ 
provided the operator $A$ does not change the total spin by 
half an integer, which is the case for all operators with finite 
cross-correlation functions with the physical currents.}
$\Theta$. In the dc limit, $\w=0$, components of
Eq.(\ref{olimit}) connecting modes of different signatures vanish.
Thus, ${M}(0)+{K}$ is block-diagonal with respect to the
time reversal signature, and consequently we can restrict our
analysis to the subspace of slow modes with the same signature as
$C_1$. However, if $C_m$ and $C_n$ have the same signature, then
$(C_m|\dot{C}_n)=0$, and thus  ${K}$  vanishes on this
restricted space. The dc conductivity therefore takes the form
\begin{equation}
\s=(M(0)^{-1})_{11}.\label{smem}
\end{equation}

\section{Central conjecture}\label{central}

To obtain a controlled approximation to the memory matrix in the
limit of small $g$, it is important to identify the relevant slow
modes of the system. For the $C_i$ we choose
quantities which are conserved by
$H_0$, $[H_0,C_i]=0$, such that $\dot{C_i}=i g [H_1,C_i]$ is
linear in the small coupling $g$.
As argued below,
we require that the singularities of correlation functions of the 
unperturbed system are exclusively due to exact conservation laws $C_i$,
i.e.~that the Drude peak appearing in Eq.~(\ref{drude}) is the only singular 
contribution.
Furthermore, we choose
$J=J_\|=C_1$ and consider only $C_i$ with the same time reversal
signature as $J$, as discussed in the previous section.

To formulate our central conjecture we introduce the following
notions. We define $M_n(\w)$ as the (exact) $n \times n$ memory
matrix obtained by setting up the memory matrix formalism for
the first $n$ slow modes $\{C_i,i=1,..,n\}$. Note that the
definitions of the relevant projectors $P$ and $Q$ also depend on this
choice, and that for any choice of $n$ one gets
$\sigma=(M_n^{-1})_{11}$. We now introduce the approximate memory
matrix $\tilde{M}_n$ motivated by the following arguments:
$\dot{C_i}$ is already {\em linear} in $g$, therefore in Eq.~(\ref{memdef}) we
approximate $L$ by $L_0$ and replace $(.|.)$
by $(.|.)_0$ as we evaluate the scalar product with respect to
$H_0$. As $L_0 |C_i)=0$ and $(C_j|\dot{C}_i)=0$ due to
time reversal symmetry, one has $L_0 Q=1$ and
$Q|\dot{C_i})=|\dot{C_i})$ and therefore the projector $Q$ does
not contribute within this approximation. We thus define the
$n \times n$ matrix $\tilde{M}_n$ by
\begin{eqnarray}
\tilde{M}_{n,ij}&=& \lim_{\w \to 0}
\left(\dot{C}_i{\Big|}\frac{i}{\w-L_0}{\Big|}\dot{C}_j\right)_0\label{memdef0}
\end{eqnarray}
Note that $\tilde{M}_n$ is a submatrix of $\tilde{M}_m$ for $m>n$
and therefore the approximate expression for the conductivity
$\sigma \approx (\tilde{M}_n^{-1})_{11}$ does depend on $n$ while
$(M_n^{-1})_{11}$ is independent of $n$. A much simpler, alternative derivation for
$\tilde{M}_1$ is given in Appendix \ref{naive}, where
the validity of this formula is also discussed.

The central conjecture of our paper is, that for small $g$ 
$(\tilde{M}_n^{-1})_{11}$ gives a lower bound to the dc conductivity,
or, more precisely,
\begin{multline} \left. \sigma\right|_{1/g^2}
=(\tilde{M}_{\infty}^{-1})_{11}\ge \dots \ge
(\tilde{M}_{n}^{-1})_{11}\ge \dots \ge \tilde{M}_{1}^{-1}.
\label{ineq}
\end{multline}
Here  $\left. \sigma\right|_{1/g^2}=(1/g^2) \lim_{g \to 0} g^2 \sigma$ denotes the
leading term $\propto 1/g^2$ in the small-$g$ expansion of $\sigma$. 
Note that $\tilde{M}_n \propto g^2$ by construction. 
$\tilde M_\infty$ is the approximate memory matrix where
{\em all}\footnote{The $C_i$ span the space of {\em all}
conservation laws, including those which do not commute with each
other.} conservation laws have been included. In some special
situations, discussed in Ref.~\onlinecite{jung.rosch:2006}, one has
$\sigma \sim 1/g^4$ and therefore  $\left. \sigma\right|_{1/g^2}=\infty$.

A special case of the inequality above is Eq.~(\ref{bound1}) in
appedix~\ref{naive}, as
the scattering rate $\tilde{\Gamma}/\chi$ may be expressed as
$\tilde{\Gamma}/\chi^2=\tilde{M}_1$.

Two steps are necessary to prove Eq.~(\ref{ineq}). The simple part
is actually the inequalities in Eq.~(\ref{ineq}). They are a
consequence of the fact that the matrices $\tilde{M}_n$ are all
positive definite and that $\tilde{M}_n$ is a  submatrix of
$\tilde{M}_m$ for $m\ge n$.  More difficult to prove is that the
first equality in (\ref{ineq}) holds. 
To show this we will need an additional assumption, namely, that the {\em regular}
part of all correlation functions (to be defined below) remains finite in the limit $g\to 0$, $\w\to 0$.
In this case, 
the perturbative expansion around $\tilde{M}_{\infty}$ in powers
of $g$ is free of singularities at finite temperature (which is not the
case for $\tilde{M}_{n < \infty}$).  This in turn implies that $\lim_{g\to 0} M_\infty/g^2 = \tilde{M}_\infty/g^2$ and therefore 
$\left. \sigma\right|_{1/g^2}
=(\tilde{M}_{\infty}^{-1})_{11}$.

Next, we present the two parts of the proof.

\subsection{Inequalities}

We start by investigating the (1,1)-component of the  inverse of
the positive definite symmetric matrix $\tilde M_{\infty}$. It is convenient to write the inverse
as\begin{eqnarray}\label{varM}
(\tilde M_{\infty}^{-1})_{11}=\max_{\bm \varphi} \frac{( \bm
\varphi^T {\bm e}_1)^2}{{\bm \varphi}^T \tilde M_\infty \bm
\varphi}
\end{eqnarray}
where ${\bm e}_1$ is the first unit vector. The same method is used to derive
 Eq.~(\ref{varBoltzmann}) in the context of the Boltzmann equation.
The maximum is
obtained for ${\bm \varphi}=\tilde M_\infty^{-1}{\bm e}_1$. By
restricting the variational space in (\ref{varM}) to the first $n$
components of $\bm \varphi$ we reproduce the submatrix
$\tilde{M}_n$ of $\tilde M_{\infty}$ and obtain
\begin{eqnarray}
(\tilde M_{\infty}^{-1})_{11} &\ge& \max_{\bm \varphi=\sum_1^m
\varphi_i {\bm e}_i} \, \frac{( \bm \varphi^T {\bm e}_1)^2}{{\bm
\varphi}^T \tilde M_\infty \bm \varphi} = (\tilde M_{m}^{-1})_{11}\nonumber\\
&\ge& \max_{\bm \varphi=\sum_1^{n<m}
\varphi_i {\bm e}_i} \, \frac{( \bm \varphi^T {\bm e}_1)^2}{{\bm
\varphi}^T \tilde M_\infty \bm \varphi} = (\tilde M_{n<m}^{-1})_{11}\nonumber
\end{eqnarray}
By choosing different values for $m$ and $n<m$, this proves all inequalities appearing in (\ref{ineq}).

\subsection{Expansion of the memory matrix}\label{expandMem}

We proceed by expanding the exact memory matrix $M_n$, where
$P_n=1-Q_n$ is a projector on the first $n$ conservation laws, in
powers of $g$. Using that $L Q_n=L_0+g L_1 Q_n$, we obtain the
geometric series
\begin{equation}
M_{n,ij}(\w)=\sum_{k=0}^\infty g^k
\left(\dot{C}_i\left|Q_n\frac{i}{\w-L_0}\left(L_1 Q_n
\frac{1}{\w-L_0}\right)^k
\right|\dot{C}_j\right).\label{MExpansion}
\end{equation}
Note that this is not a full expansion in $g$, as the scalar product (\ref{scalarProduct}) is defined with respect to the full Hamiltonian $H=H_0+g
H_1$. We will turn to the discussion of the remaining
$g$-dependence later.

In general, one can expand 
\[ L_1=\sum_{m,n}\lambda_{mn} |A_m)(A_n|\] 
in terms of some basis $A_m$ in the space of operators.
Therefore Eq.~(\ref{MExpansion}) can be written as a sum over products
of terms with the general structure
\be
\left(A\left|Q_n\frac{1}{\w-L_0}\right|B\right).\label{c1}
\ee
In the following we would like to argue that such an expansion is
regular for $n=\infty$ if {\em all} conservation laws have been
included in the definition of $Q$. 
As argued in Appendix~\ref{naive}, we have to investigate whether the series coefficients in
Eq.~(\ref{MExpansion}) diverge for $\w \to 0$.
The basis of our argument is the
following: as $Q_\infty$ projects the dynamics to the space
perpendicular to all of the conservation laws, the associated singularities
 are absent 
in Eq.~(\ref{c1}) and therefore the expansion of $M_\infty$ is regular.

To show this more formally, we split up $B=B_\parallel+B_\perp$
in (\ref{c1})  into a component parallel and one perpendicular to
the  space of {\em all} conserved quantities, $|B_\|)=P_\infty
|B)$. With this notation, the action of $L_0$ becomes more
transparent:
\begin{equation}
\frac{1}{\w-L_0}|B)=\frac{1}{\w}|B_\parallel)+\frac{1}{\w-L_0}|B_\perp).\label{c2}
\end{equation}
As we assume that all divergencies can be traced back to 
the conservation laws, we take the second term to be 
regular. It is only the first term which leads in
Eq.~(\ref{c1}) to a divergence for $\w \to 0$, provided that
$(A|Q_n|B_\|)$ is finite. If we consider the perturbative
expansion of $M_{n<\infty}$, where $P_n=1-Q_n$ projects only to a
subset of conserved quantities, then finite contributions of the
form $(A|Q_n|B_\|)$ exist and the perturbative series in $g$ will
be singular (see also Appendix \ref{naive}). Considering
$M_\infty$, however, $Q_\infty$ projects out {\em all}
conservation laws and therefore by construction $Q_\infty
|B_\|)=Q_\infty P_\infty |B)=0$. Thus the first term in (\ref{c2})
does not contribute in (\ref{c1}) for $n=\infty$ and the expansion
(\ref{MExpansion}) of $M_\infty$ is therefore regular.

The only remaining part of our argument is to show that in the
limit $g \to 0$ one can safely replace $(.|.)$ by $(.|.)_0$. Here
it is useful to realize that $(A|B)$ can be interpreted as a
(generalized) static susceptibility. In the absence of a phase
transition and at finite temperatures, susceptibilities are
smooth, non-singular functions of the coupling constants and
therefore we do not expect any further singularities from this
step. If we define a phase transition by a singularity in some
generalized susceptibility, then the statement that
susceptibilities are regular in the absence of phase transitions
even becomes a mere tautology.

Combining all arguments, the expansion (\ref{MExpansion}) of
$M_\infty(\w \to 0)$
is regular, and using $(\dot{C_i}|Q_\infty=(\dot{C_i}|$ 
[see discussion before Eq.~(\ref{memdef0})]
its leading term, $k=0$
is given by $\tilde{M}_\infty$.
We therefore have shown the
missing first equality of our central conjecture~(\ref{ineq}).

\section{Discussion}\label{discussion}

In this paper we have established that in
the limit of small perturbations, $H=H_0+g H_1$,  lower bounds to
dc conductivities may be calculated for situations where the
conductivity is infinite for $g=0$. In the opposite case, when the
conductivity is finite for $g=0$, one can use naive perturbation
theory to calculate small corrections to $\sigma$ without further
complications.

The relevant lower bounds are directly obtained from the memory
matrix formalism. Typically\cite{gotze.wolfle:1972,giamarchi:1991,%
rosch.andrei:2000} one has to
evaluate a small number of correlation functions and to invert small
matrices. The quality of the lower bounds depends decisively on
whether one has been able to identify the `slowest' modes in the
system.

There are many possible applications for the results presented in
this paper. The mostly considered situation is the case where
$H_0$ describes a non-interacting system\cite{gotze.wolfle:1972}. 
For situations where the Boltzmann
equation can be applied, it has been pointed out a long time ago
by Belitz \cite{belitz:1984} that there is a one-to-one relation of
the memory matrix calculation to a certain variational Ansatz to
the Boltzmann equation, see Eq.~(\ref{varBoltzmann}). In this paper we
were able to generalize this result to cases where a Boltzmann description is not
possible. For example, if $H_0$ is the Hamiltonian of a Luttinger
liquid, i.e.~a non-interacting bosonic system, then typical
perturbations are of the form $\cos \phi$ for which a simple
transport theory in the spirit of a Boltzmann or vertex equation
does not exist to our knowledge.

Another class of applications are systems where $H_0$ describes an
{\em interacting} system, e.g.~an integrable one-dimensional
model\cite{jung.rosch:2006} or some non-trivial quantum-field
theory\cite{boulat-2006}. In these cases it can become difficult to
calculate the memory matrix and one has to resort to use either
numerical\cite{jung.rosch:2006} or field-theoretical methods\cite{boulat-2006}
to obtain the relevant correlation functions.

An important special case are situations where $H_0$ is
characterized by a {\em single} conserved current with the proper
symmetries, i.e.~with overlap to the (heat-, spin- or charge-)
current $J$.  For example, in a non-trivial continuum field theory
$H_0$, interactions lead to the decay of all modes with exception
of the momentum $P$. In this case the momentum relaxation and
therefore the conductivity at finite $T$ is determined by small
perturbations $g H_1$ like disorder or Umklapp scattering which
are present in almost any realistic system. As
$\tilde{M}_\infty=\tilde{M}_1$ in this case, our
results suggest that for small $g$  the conductivity is {\em
exactly} determined by the momentum relaxation rate
$\tilde{M}_{PP}=\lim_{\w \to 0} i(\dot{P}|(\w-L_0)^{-1}|\dot{P})$,
\be
\sigma = \frac{\chi_{PJ}^2}{\tilde{M}_{PP}} \quad {\rm for} \ g \to 0.  \label{oneCL}
\ee
Here we used that $J_\|=P (P|J)/(P|P)$ with
$\chi_{PJ}=(P|J)$ and we have restored all factors which arise if
the normalization condition (\ref{normC}) is not used. In
Appendix~\ref{singleSlowMode}, we check numerically that this statement is
valid for a realistic example within the Boltzmann equation
approach.

A number of assumptions entered our arguments. The strongest
one is the restriction that all relevant singularities arise
from exact conservation laws of $H_0$. We assumed that
the regular parts of correlation functions  are finite for $\w=0$. 
There are two distinct scenarios in which this assumption does not hold.
First, in the limit $T\to 0$, often scattering rates vanish which can lead to diverencies of the nominally regular parts  of correlation functions. Furthermore, at $T=0$ even infinitesimally small perturbations can induce phase transitions -- again a situation where our arguments fail. Therefore our results
are not applicable at $T=0$. 
Second, finite temperature transport  may be plagued by additional 
divergencies for $\w\to 0$ not 
captured by the Drude weight. 
In some special models, for instance,
transport is singular even in the {\em absence} of exactly
conserved quantities (e.g. non-interacting phonons in a disordered
crystal\cite{ziman:1960}). In all cases known to us, these divergencies
can be traced back to the presence of some slow modes in the
system (e.g. phonons with very low momentum). While we have not
kept track of such divergencies in our arguments, we nevertheless
believe that they do {\em not} invalidate our main inequality
(\ref{ineq}) as further slow modes not captured by exact
conservation laws will only increase the conductivity. It is,
however, likely that the {\em equality} (\ref{oneCL}) is not
valid for such situations. In Appendix 
\ref{singleSlowMode} we analyze in some detail within the Boltzmann equation formalism 
under which conditions (\ref{oneCL}) holds.
As an aside, we note that the singular
heat transport of non-interacting disordered phonons, mentioned
above, is well described within our formalism if we model the
clean system by $H_0$ and the disorder by $H_1$, see the extensive
discussion by Ziman\cite{ziman:1960} within the variational approach
which can be directly translated to the memory matrix language,
see Ref.~[\onlinecite{belitz:1984}].

It would be interesting to generalize our results to cases where
time reversal symmetry is broken, e.g.~by an external magnetic
field. As time reversal invariance entered nontrivially in our
arguments, this seems not to be simple. We nevertheless do 
not see any physical
reason why the inequality should not be valid in this case, too. One
example where no problems arise are spin chains in a uniform
magnetic field\cite{sologubenko:2007} where one can map the
field to a chemical potential using a Jordan-Wigner
transformation. Then one can directly apply our results to the
time reversal invariant system of Jordan-Wigner fermions.

\acknowledgements
We thank N. Andrei, E.~Shimshoni, P. W\"olfle and X.~Zotos
for useful discussions. This work was partly supported by the
Deutsche Forschungsgemeinschaft through SFB 608 and the German
Israeli Foundation.

\appendix
\section{Drude weight and Mazur inequality}\label{appA}

In this appendix we clarify the connection between the Drude 
weight and the Mazur inequality, mentioned in the introduction.

The Drude weight $D$ is the singular part of the conductivity at zero
frequency, $\Re\s(\w)=\pi D \delta(\w)+\s_{\rm reg}(\w)$. It can be
calculated from the relation $$D= \lim_{\w\ra0}\w \;\Im\s(\w).$$
It has been introduced by Kohn \cite{kohn:1964} as a measure of
ballistic transport, indicated by $D>0$.

Using Kubo formulas, conductivities can be expressed in terms of 
the dynamic current susceptibilities\cite{kadanoff.martin:1963} $\Pi(z)$ using
\be
\s(z)=-\frac{1}{i z}\left( \Pi^T - \Pi(z) \right),\label{kubo}
\ee
where $\Pi(z)$ is the current response function
\begin{eqnarray} 
\Pi(z)&=&\frac{i}{V}\int_0^\infty dt e^{izt}\langle[J(t),J(0)]\rangle\\ 
\Pi^T&=&\int \frac{d\w}{\pi}\frac{\Pi''(\w)}{\w}\label{chiT}. 
\end{eqnarray} 
and $\Pi^T$ is a current susceptibility. The conductivity
may be calculated by setting $\s(\w)=\s(z=\w+i0)$. Relation
(\ref{chiT}) is a well known sum rule and for all {\em regular} 
correlation functions one has $\Pi^T=\Pi(0)$.
In the presence of a singular contribution to $\sigma(\w)$, one easily
identifies the Drude weight with the expression $\Pi^T-\Pi(0)$.
For this difference Mazur\cite{mazur:1969,
suzuki:1971} derived a lower bound. Furthermore, Suzuki\cite{suzuki:1971} has shown,
that $\Pi^T-\Pi(0)$ may be expressed as a sum over all constants of the motion
$C_i$ present in the system\footnote{More precisely,  $\{C_i\}$ is
taken to be a basis of the space of operators with energy-diagonal
entries only, chosen to be orthogonal in the sense that $\langle C_i
C_j \rangle\propto \delta_{ij}$.},
\begin{equation}
D=\Pi^T-\Pi(0)=\frac{\beta}{V}\sum_{n=0}^\infty \frac{\langle C_j J \rangle^2}{\langle
C_j^2 \rangle}.
\end{equation} 
Thus, the Drude weight is intimately connected to the presence of
conservation laws:  only
components of the current perpendicular to all conservation laws decay and any
conservation law with a component parallel to the current 
(i.e. with a finite cross-correlation $\langle C_j J \rangle$)
leads to a finite Drude weight and thus ballistic transport.
The relation between the Drude weight and Mazur's
inequality has been first pointed out by Zotos\cite{zotos:1997}.

\section{Perturbation theory for $1/\sigma$}\label{naive}
Let us give an example of a naive perturbative derivation 
(see also Ref.~[\onlinecite{jung.rosch:2006}]) to gain
some insight about what problems can turn up in a 
perturbative derivation as the one presented in this work. According to our
assumptions, the conductivity is diverging for $g \to 0$  and
therefore it is useful to consider the scattering rate
$\Gamma(\w)/\chi$ (with the current susceptibility $\chi$) defined by
\begin{equation}
\s(\w)=\frac{\chi}{\Gamma(\w)/\chi-i \w}.\label{Gammadef}
\end{equation}
If $J$ is conserved for $g=0$ (i.e. for $J=J_\|$, see above), the
scattering rate vanishes, $\Gamma(\w)=0$, for $g=0$, which results
in a finite Drude weight. A perturbation around this singular
point results in a finite $\Gamma(\w)$. In the limit $g\to 0$ we
can expand (\ref{Gammadef}) for any {\em finite} frequency $\w$ in
$\Gamma$ to obtain
\begin{equation}
\w^2 \Re\s(\w) = \Re
\Gamma(\w)+\mathcal{O}(\Gamma^2/\w).\label{Gammaexp}
\end{equation}
We can read this as an equation for the leading order contribution
to $\Gamma(\w)$, which now is expressed through the Kubo formula
for the conductivity. By partially integrating twice in time we
can write $\Gamma(\w)= \tilde{\Gamma}(\w)+\mathcal{O}(g^3)$ with
\be
\Re\tilde{\Gamma}(\w)=\Re \frac{1}{z}\frac{1}{V}\int_0^\infty dt
e^{i z t}\langle
 [\dot{J}(t),\dot{J}(0)]\rangle_0\Big{|}_{z=\w+i0}, \label{gammaexp2}
\ee
where $\dot{J}=i [H,J]=i g [H_1,J]$ is linear in $g$ and therefore
the expectation value $\langle ...\rangle_0$  can be evaluated
with respect to $H_0$ (which may describe an interacting system).
Thus we have expressed the scattering rate via a simple
correlation function of the time derivative of the current.

To determine the dc conductivity one is interested in the limit
$\w\to 0$ and it is tempting to set $\w=0$ in
Eq.~(\ref{gammaexp2}). We have, however, derived
Eq.~(\ref{gammaexp2}) in the limit $g \to 0$ at finite $\omega$
and {\em not} in the limit $\w \to 0$ at finite $g$. The series
Eq.~(\ref{Gammaexp}) is well defined for finite $\w\neq 0$ only and
in the limit $\w\to 0$ the series shows singularities to
arbitrarily high orders in $1/\w$.

At first sight this makes Eq.~(\ref{gammaexp2}) useless for
calculating the dc conductivity. One of the main results of this
paper is that, nevertheless, $\tilde \Gamma(\w=0)$ can be used to
obtain a lower bound to the dc conductivity
\be
\sigma(\w=0) \ge \frac{\chi^2}{\tilde{\Gamma}(0)} \quad 
{\rm for } \ g \to 0. \label{bound1}
\ee

\section{Single Slow Mode} \label{singleSlowMode}
In this appendix we check whether in the presence of a 
{\em single} conservation law with finite cross correlations 
with the current the inequality (\ref{ineq}) can be replaced by
the equality (\ref{oneCL}). This requires us to compare the 
true conductivity, which in general is hard to
determine, to the result given by
$\tilde{M}_1$. Thus we restrict ourselves to the discussion of models
for which a Boltzmann equation can be formulated and the expression
for the conductivity can be calculated at least numerically.
In the following we first
show numerically that the equality (\ref{oneCL}) holds for a realistic model.
In a second step we discuss the precise regularity 
requirement of the scattering matrix such that Eq.~(\ref{oneCL}) holds.

To simplify numerics, we consider a simple one-dimensional 
Boltzmann equation of interacting and weakly disordered
Fermions. Clearly, the Boltzmann approach breaks down 
close to the Fermi surface due to singularities associated 
with the formation of a Luttinger liquid, but in the present context
we are not interested in this physics as we only want to 
investigate properties of the Boltzmann equation.
To avoid the restrictions associated with momentum and energy 
conservation in one dimension we consider a dispersion with
two minima and four Fermi points,
\begin{equation}
\epsilon_k=-\frac{k^2}{2}+\frac{k^4}{4}+\frac{1}{10}.\label{disp}
\end{equation}
The Boltzmann equation reads 
\begin{align} v_k
\frac{d f^0_k}{d \epsilon_k} E &=\sum_{k'qq'} S_{kk'}^{qq'} \bigl[
f_k f_{k'} (1-f_q)(1-f_{q'}) \nonumber \\ 
&\qquad \qquad -f_q f_{q'} (1-f_k)(1-f_{k'}) \bigr] \nonumber \\ 
&+ g^2 \sum_{k'}  \delta(\epsilon_k-\epsilon_{k'})\bigl[ f_k(1-f_{k'})-
f_{k'}(1-f_{k})\bigr] \nonumber \\ 
&= \sum_{k'} W_{kk'} \Phi_{k'} \label{lin} 
\end{align} 
where the inelastic scattering term $S_{kk'}^{qq'}=\delta(\epsilon_k+\epsilon_{k'}-
\epsilon_q-\epsilon_{q'}) \delta(k+{k'}-q-{q'})$ 
conserves both energy and momentum. In the last line we have linearized
the right hand side using the definitions of the introductory chapter. The
velocity $v_k$ is given by $v_k=\frac{d}{d k}
\epsilon_k$. The scattering matrix splits up into an
\emph{interaction} component and a \emph{disorder} component, $W_{k
k'}=W_{k k'}^{0}+g^2 W_{k k'}^{1}$. As we do not consider 
Umklapp scattering, $W_{k k'}^{0}$ conserves momentum, 
$\sum_{k'} W^0_{k k'} k'=0$, and one expects that momentum relaxation will 
determine the conductivity for small $g$.

For the numerical calculation we discretize momentum in the interval
$[-\pi/2,\pi/2]$, $k_n=n \delta k= n \pi/N$ with integer $n$.  (At the
boundaries the energy is already too high to play any role in
transport.) The delta function arising from energy conservation is
replaced by a gaussian of width $\delta$. The proper thermodynamic
limit can for example be obtained by choosing $\delta = 0.3/\sqrt{N}$.
The numerics shows small finite size effects. %
\begin{figure} \begin{center} 
\includegraphics[width=.98\linewidth,clip=]{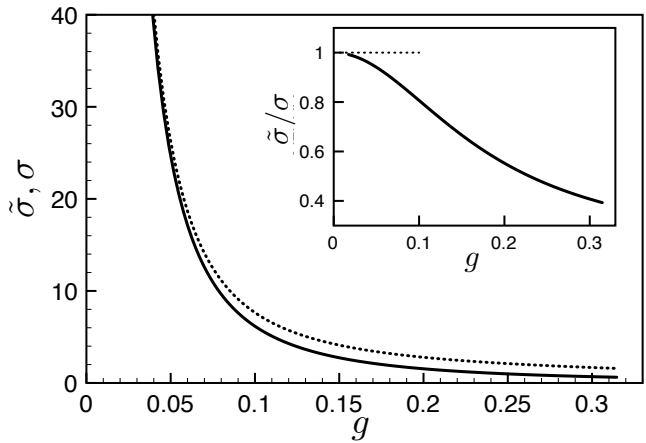} 
\end{center} \caption{\label{boltzfig}
Comparison of the result of a single mode memory matrix calculation
(solid line), Eq. (\ref{mmB}),  to the full numerical solution of the Boltzmann
equation (dotted line) for
$T=0.05$ and $N=500$. The memory matrix is always a lower bound to the
Boltzmann result and converges towards it as the disorder strength
$g$ is reduced, as shown in the inset (ratio of the single mode
approximation to the Boltzmann result). } \end{figure} %

In Fig.~\ref{boltzfig} we compare the numerical 
solution of the Boltzmann equation
to the single mode memory matrix calculation or, 
equivalently\cite{belitz:1984}, to the variational bound obtained
 by setting $\Phi_k=k$ in Eq.~(\ref{varBoltzmann})
 \begin{equation}\label{mmB}
 \tilde{\sigma}=\frac{ \left( \sum_k {v}^i_{k} k
\frac{d f^0}{d \epsilon_{k}}\right)^2}{\sum_{ k, k'}
k W_{{k}{k}'} k'}=\frac{ \left( \sum_{k} {v}^i_{{k}} k
\frac{d f^0}{d \epsilon_{k}}\right)^2}{g^2 \sum_{ k, k'}
k W^1_{{k}{k}'} k'}.
 \end{equation} As can be seen from the inset, in the limit of small $g$
 one obtains the exact value for the
conductivity, which is what we intended to demonstrate.

 Next we turn
to an analysis of regularity conditions which have to be met in general by the
scattering matrix $W_{k k'}$ such that convergence is guaranteed in
the limit $g\to 0$.
According to the assumptions of this appendix, for $g=0$ the
variational form of the Boltzmann equation (\ref{varBoltzmann}) has a
unique solution $\bar{\Phi}_k$ (up to a multiplicative constant), with
$F(\bar{\Phi}_k)=\infty$, $\sum_{k'} W^0_{k k'}\bar{\Phi}_{k'}=0$ and
 $\sum_{k} v_k  \bar{\Phi}_{k} df^0/d \epsilon_k>0$.

In the presence of a finite, but small $g$ we write the solution of the Boltzmann equation
as $\Phi=\bar{\Phi}+ \Phi^\perp$, where $\Phi^\perp$ has no
component parallel to $\bar{\Phi}$ (i.e.~$\sum_k \bar{\Phi}_k
\Phi^\perp_k df^0/d \epsilon_k=0$ ).
On the basis of the two inequalities
\begin{eqnarray}
F[\bar{\Phi}] &\le& F[\Phi] \\
\Phi W \Phi &=& \bar{\Phi} g^2 W^1 \bar{\Phi}
+\Phi_\perp W \Phi_\perp \ge \bar{\Phi} g^2 W^1 \bar{\Phi}
\end{eqnarray}
one concludes that Eq.~(\ref{oneCL}) is valid, i.e.~that
\[
\lim_{g \to 0} \frac{F[\bar{\Phi}]}{F[\Phi]} = 1
\]
under the condition that
\[
\lim_{g\to 0} \sum_k v_k \Phi_k \frac{d f^0}{d \epsilon_k} 
= \sum_k v_k \bar{\Phi}_k \frac{d f^0}{d \epsilon_k}
\]
or, equivalently,
\be
\lim_{g\to 0} \sum_k v_k \Phi_{\perp,k} \frac{d f^0}{d \epsilon_k} = 0. \label{limEq}
\ee
We therefore have to check whether $\Phi_\perp$ becomes small in the limit of small $g$.

Expanding the saddlepoint equation for (\ref{varBoltzmann})
we obtain
\begin{align} \sum_{k'} W_{k
k'}^0 \Phi^\perp_{k'} = v_{k} \frac{d f^0}{d \epsilon_{k}}
\frac{\sum_{k' k''} \bar{\Phi}_{k'}  g^2 W_{k' k''}^1
\bar{\Phi}_{k''}}{ \sum_{k'} v_{k'} \frac{d f^0}{d \epsilon_{k'}}
\bar{\Phi}_{k'} } \nonumber \\
- \sum_{k'}  g^2 W_{k k'}^1 \bar{\Phi}_{k'}
+\mathcal{O}(g^2 W_1 \Phi_\perp, \Phi_\perp W_0 \Phi_\perp) \nonumber
\end{align}
As by definition $\Phi^\perp$ has no component parallel to $\bar{\Phi}$,
we can insert the projector $Q$ which projects out the conservation
law in front of  $\Phi^\perp_k$ on the left hand side.
We therefore conclude that {\em if} the inverse of $W^0 Q$ 
exists, then $\Phi_\perp$ is of order $g^2$,
Eq.~(\ref{limEq}) is valid and therefore also Eq.~(\ref{oneCL}).
In our numerical examples these conditions are all met.

Under what conditions can one expect that Eq.~(\ref{limEq}) is not 
valid? Within the assumptions of this appendix we
have excluded the presence of other zero modes of $W^0$ (i.e. conservation laws)
with finite overlap with the current. But it may happen that $W^0$ 
has many eigenvalues which are arbitrarily small
such that the sum in Eq.~(\ref{limEq}) diverges. 
In such a situation the presence of slow modes which cannot be
identified with conservation laws of the unperturbed system 
invalidates Eq.~(\ref{oneCL}).



\begin{thebibliography}{33}
\expandafter\ifx\csname natexlab\endcsname\relax\def\natexlab#1{#1}\fi
\expandafter\ifx\csname bibnamefont\endcsname\relax
  \def\bibnamefont#1{#1}\fi
\expandafter\ifx\csname bibfnamefont\endcsname\relax
  \def\bibfnamefont#1{#1}\fi
\expandafter\ifx\csname citenamefont\endcsname\relax
  \def\citenamefont#1{#1}\fi
\expandafter\ifx\csname url\endcsname\relax
  \def\url#1{\texttt{#1}}\fi
\expandafter\ifx\csname urlprefix\endcsname\relax\def\urlprefix{URL }\fi
\providecommand{\bibinfo}[2]{#2}
\providecommand{\eprint}[2][]{\url{#2}}

\bibitem[{\citenamefont{Zotos and Prelovsek}(1996)}]{zotos.prelovsek:1996}
\bibinfo{author}{\bibfnamefont{X.}~\bibnamefont{Zotos}} \bibnamefont{and}
  \bibinfo{author}{\bibfnamefont{P.}~\bibnamefont{Prelovsek}}, \bibinfo{journal}{\prb}
  \textbf{\bibinfo{volume}{53}}, \bibinfo{pages}{983} (\bibinfo{year}{1996}).

\bibitem[{\citenamefont{Fabricius and McCoy}(1998)}]{fabricius.mccoy:1998}
\bibinfo{author}{\bibfnamefont{K.}~\bibnamefont{Fabricius}} \bibnamefont{and}
  \bibinfo{author}{\bibfnamefont{B.~M.} \bibnamefont{McCoy}},
  \bibinfo{journal}{\prb} \textbf{\bibinfo{volume}{57}},
  \bibinfo{pages}{8340} (\bibinfo{year}{1998}).

\bibitem[{\citenamefont{Narozhny et~al.}(1998)\citenamefont{Narozhny, Millis,
  and Andrei}}]{narozhny:millis.andrei:1998}
\bibinfo{author}{\bibfnamefont{B.~N.} \bibnamefont{Narozhny}},
  \bibinfo{author}{\bibfnamefont{A.~J.} \bibnamefont{Millis}},
  \bibnamefont{and} \bibinfo{author}{\bibfnamefont{N.}~\bibnamefont{Andrei}},
  \bibinfo{journal}{\prb} \textbf{\bibinfo{volume}{58}},
  \bibinfo{pages}{R2921} (\bibinfo{year}{1998}).

\bibitem[{\citenamefont{Zotos and Prelovsek}(2003)}]{zotos.prelovsek:2003}
\bibinfo{author}{\bibfnamefont{X.}~\bibnamefont{Zotos}} \bibnamefont{and}
  \bibinfo{author}{\bibfnamefont{P.}~\bibnamefont{Prelovsek}}, 
 \eprint{e-print arXiv:cond-mat/0304630} (\bibinfo{year}{2003}).
 
\bibitem[{\citenamefont{{Heidrich-Meisner}
  et~al.}(2005{\natexlab{a}})\citenamefont{{Heidrich-Meisner}, {Honecker},
  {Cabra}, and {Brenig}}}]{heidrich-meissner:2005-b}
\bibinfo{author}{\bibfnamefont{F.}~\bibnamefont{{Heidrich-Meisner}}},
  \bibinfo{author}{\bibfnamefont{A.}~\bibnamefont{{Honecker}}},
  \bibinfo{author}{\bibfnamefont{D.~C.} \bibnamefont{{Cabra}}},
  \bibnamefont{and} \bibinfo{author}{\bibfnamefont{W.}~\bibnamefont{{Brenig}}},
  \bibinfo{journal}{Physica B} \textbf{\bibinfo{volume}{359}},
  \bibinfo{pages}{1394} (\bibinfo{year}{2005}{\natexlab{a}}).

\bibitem[{\citenamefont{Jung et~al.}(2006)\citenamefont{Jung, Helmes, and
  Rosch}}]{jung.rosch:2006}
\bibinfo{author}{\bibfnamefont{P.}~\bibnamefont{Jung}},
  \bibinfo{author}{\bibfnamefont{R.~W.} \bibnamefont{Helmes}},
  \bibnamefont{and} \bibinfo{author}{\bibfnamefont{A.}~\bibnamefont{Rosch}},
  \bibinfo{journal}{\prl} \textbf{\bibinfo{volume}{96}},
  \bibinfo{pages}{067202} (\bibinfo{year}{2006}).

\bibitem[{\citenamefont{{Schollw{\"o}ck}}(2005)}]{schollwoeck:2005}
\bibinfo{author}{\bibfnamefont{U.}~\bibnamefont{{Schollw{\"o}ck}}},
  \bibinfo{journal}{Rev.~Mod.~Phys.} \textbf{\bibinfo{volume}{77}},
  \bibinfo{pages}{259} (\bibinfo{year}{2005}).

\bibitem[{\citenamefont{Ziman}(1960)}]{ziman:1960}
\bibinfo{author}{\bibfnamefont{J.}~\bibnamefont{Ziman}},
  \emph{\bibinfo{title}{Electrons and Phonons: The theory of transport
  phenomena in solids}} (\bibinfo{publisher}{Oxford University Press},
  \bibinfo{year}{1960}).

\bibitem[{\citenamefont{Kohler}(1948)}]{kohler:1948}
\bibinfo{author}{\bibfnamefont{M.}~\bibnamefont{Kohler}}, \bibinfo{journal}{Z.
  Phys.} \textbf{\bibinfo{volume}{124}}, \bibinfo{pages}{772}
  (\bibinfo{year}{1948}).

\bibitem[{\citenamefont{Kohler}(1949)}]{kohler:1949}
\bibinfo{author}{\bibfnamefont{M.}~\bibnamefont{Kohler}}, \bibinfo{journal}{Z.
  Phys.} \textbf{\bibinfo{volume}{125}}, \bibinfo{pages}{679}
  (\bibinfo{year}{1949}).

\bibitem[{\citenamefont{{Sondheimer}}(1950)}]{sondheimer:1950}
\bibinfo{author}{\bibfnamefont{E.~H.} \bibnamefont{{Sondheimer}}},
  \bibinfo{journal}{Proc.~R.~Soc.~London, Ser.~A}
  \textbf{\bibinfo{volume}{203}}, \bibinfo{pages}{75} (\bibinfo{year}{1950}).

\bibitem[{\citenamefont{{Mazur}}(1969)}]{mazur:1969}
\bibinfo{author}{\bibfnamefont{P.}~\bibnamefont{{Mazur}}},
  \bibinfo{journal}{Physica (Amsterdam)} \textbf{\bibinfo{volume}{43}},
  \bibinfo{pages}{533} (\bibinfo{year}{1969}).

\bibitem[{\citenamefont{{Suzuki}}(1971)}]{suzuki:1971}
\bibinfo{author}{\bibfnamefont{M.}~\bibnamefont{{Suzuki}}},
  \bibinfo{journal}{Physica (Amsterdam)} \textbf{\bibinfo{volume}{51}},
  \bibinfo{pages}{277} (\bibinfo{year}{1971}).

\bibitem[{\citenamefont{Zotos et~al.}(1997)\citenamefont{Zotos, Naef, and
  Prelovsek}}]{zotos:1997}
\bibinfo{author}{\bibfnamefont{X.}~\bibnamefont{Zotos}},
  \bibinfo{author}{\bibfnamefont{F.}~\bibnamefont{Naef}}, \bibnamefont{and}
  \bibinfo{author}{\bibfnamefont{P.}~\bibnamefont{Prelovsek}},
  \bibinfo{journal}{\prb} \textbf{\bibinfo{volume}{55}},
  \bibinfo{pages}{11029} (\bibinfo{year}{1997}).

\bibitem[{\citenamefont{Fujimoto and Kawakami}(2003)}]{fujimoto.kawakami:2003}
\bibinfo{author}{\bibfnamefont{S.}~\bibnamefont{Fujimoto}} \bibnamefont{and}
  \bibinfo{author}{\bibfnamefont{N.}~\bibnamefont{Kawakami}},
  \bibinfo{journal}{\prl} \textbf{\bibinfo{volume}{90}},
  \bibinfo{pages}{197202} (\bibinfo{year}{2003}).

\bibitem[{\citenamefont{{Heidrich-Meisner}
  et~al.}(2005{\natexlab{b}})\citenamefont{{Heidrich-Meisner}, {Honecker}, and
  {Brenig}}}]{heidrich-meissner:2005}
\bibinfo{author}{\bibfnamefont{F.}~\bibnamefont{{Heidrich-Meisner}}},
  \bibinfo{author}{\bibfnamefont{A.}~\bibnamefont{{Honecker}}},
  \bibnamefont{and} \bibinfo{author}{\bibfnamefont{W.}~\bibnamefont{{Brenig}}},
  \bibinfo{journal}{\prb} \textbf{\bibinfo{volume}{71}},
  \bibinfo{pages}{184415} (\bibinfo{year}{2005}{\natexlab{b}}).

\bibitem[{\citenamefont{{Sakai}}(2005)}]{sakai:2005}
\bibinfo{author}{\bibfnamefont{K.}~\bibnamefont{{Sakai}}},
  \bibinfo{journal}{Physica E (Amsterdam)}
  \textbf{\bibinfo{volume}{29}}, \bibinfo{pages}{664} (\bibinfo{year}{2005}).

\bibitem[{\citenamefont{Shimshoni et~al.}(2003)\citenamefont{Shimshoni, Andrei,
  and Rosch}}]{shimshoni:2003}
\bibinfo{author}{\bibfnamefont{E.}~\bibnamefont{Shimshoni}},
  \bibinfo{author}{\bibfnamefont{N.}~\bibnamefont{Andrei}}, \bibnamefont{and}
  \bibinfo{author}{\bibfnamefont{A.}~\bibnamefont{Rosch}},
  \bibinfo{journal}{\prb} \textbf{\bibinfo{volume}{68}},
  \bibinfo{pages}{104401} (\bibinfo{year}{2003}).
  
\bibitem[{\citenamefont{{Heidrich-Meisner}
  et~al.}(2002)\citenamefont{{Heidrich-Meisner}, {Honecker}, {Cabra}, and
  {Brenig}}}]{heidrich-meissner:2002}
\bibinfo{author}{\bibfnamefont{F.}~\bibnamefont{{Heidrich-Meisner}}},
  \bibinfo{author}{\bibfnamefont{A.}~\bibnamefont{{Honecker}}},
  \bibinfo{author}{\bibfnamefont{D.~C.} \bibnamefont{{Cabra}}},
  \bibnamefont{and} \bibinfo{author}{\bibfnamefont{W.}~\bibnamefont{{Brenig}}},
  \bibinfo{journal}{\prb} \textbf{\bibinfo{volume}{66}},
  \bibinfo{pages}{140406(R)} (\bibinfo{year}{2002}).

\bibitem[{\citenamefont{Rozhkov and
  Chernyshev}(2005)}]{rozhkov.chernyshev:2005}
\bibinfo{author}{\bibfnamefont{A.~V.} \bibnamefont{Rozhkov}} \bibnamefont{and}
  \bibinfo{author}{\bibfnamefont{A.~L.} \bibnamefont{Chernyshev}},
  \bibinfo{journal}{\prl} \textbf{\bibinfo{volume}{94}},
  \bibinfo{pages}{087201} (\bibinfo{year}{2005}).
  
\bibitem[{\citenamefont{{Sologubenko} et~al.}(2000)\citenamefont{{Sologubenko},
  {Felder}, {Giann{\`o}}, {Ott}, {Vietkine}, and
  {Revcolevschi}}}]{sologubenko:2000}
\bibinfo{author}{\bibfnamefont{A.~V.} \bibnamefont{{Sologubenko}}},
  \bibinfo{author}{\bibfnamefont{E.}~\bibnamefont{{Felder}}},
  \bibinfo{author}{\bibfnamefont{K.}~\bibnamefont{{Giann{\`o}}}},
  \bibinfo{author}{\bibfnamefont{H.~R.} \bibnamefont{{Ott}}},
  \bibinfo{author}{\bibfnamefont{A.}~\bibnamefont{{Vietkine}}},
  \bibnamefont{and}
  \bibinfo{author}{\bibfnamefont{A.}~\bibnamefont{{Revcolevschi}}},
  \bibinfo{journal}{\prb} \textbf{\bibinfo{volume}{62}}, \bibinfo{pages}{R6108}
  (\bibinfo{year}{2000}).

\bibitem[{\citenamefont{Hess et~al.}(2007)\citenamefont{Hess, ElHaes, Waske,
  Buchner, Sekar, Krabbes, Heidrich-Meisner, and Brenig}}]{hess:2007}
\bibinfo{author}{\bibfnamefont{C.}~\bibnamefont{Hess}},
  \bibinfo{author}{\bibfnamefont{H.}~\bibnamefont{El Haes}},
  \bibinfo{author}{\bibfnamefont{A.}~\bibnamefont{Waske}},
  \bibinfo{author}{\bibfnamefont{B.}~\bibnamefont{Buchner}},
  \bibinfo{author}{\bibfnamefont{C.}~\bibnamefont{Sekar}},
  \bibinfo{author}{\bibfnamefont{G.}~\bibnamefont{Krabbes}},
  \bibinfo{author}{\bibfnamefont{F.}~\bibnamefont{Heidrich-Meisner}},
  \bibnamefont{and} \bibinfo{author}{\bibfnamefont{W.}~\bibnamefont{Brenig}},
  \bibinfo{journal}{\prl} \textbf{\bibinfo{volume}{98}},
  \bibinfo{eid}{027201} (\bibinfo{year}{2007}).
  
\bibitem[{\citenamefont{{Forster}}(1975)}]{forster:1975}
  \emph{\bibinfo{title}{{Hydrodynamic Fluctuations, Broken Symmetry, and
  Correlation Functions}}}, 
  \bibinfo{editor}{edited by \bibfnamefont{D.}~\bibnamefont{{Forster}}}
  (Perseus, New York, \bibinfo{year}{1975}).

\bibitem[{\citenamefont{{Mori}}(1965)}]{mori:1965}
\bibinfo{author}{\bibfnamefont{H.}~\bibnamefont{{Mori}}},
  \bibinfo{journal}{Prog.~Theor.~Phys.}
  \textbf{\bibinfo{volume}{33}}, \bibinfo{pages}{423} (\bibinfo{year}{1965}).

\bibitem[{\citenamefont{Zwanzig}(1960)}]{zwanzig:1960}
\bibinfo{author}{\bibfnamefont{R.}~\bibnamefont{Zwanzig}}, in
  \emph{\bibinfo{booktitle}{Lectures in Theoretical Physics,}} edited by W.~E.~
  Brittin, B.~W.~Downs and J.~Downs (Interscience, New York,
  1961), Vol.~III; \bibinfo{editor}{\bibfnamefont{J.~Chem.}
  \bibnamefont{Phys.}} \bibinfo{volume}{{\bf 33}}, \bibinfo{pages}{1338} (\bibinfo{year}{1960}).

\bibitem[{\citenamefont{{G{\"o}tze} and
  {W{\"o}lfle}}(1972)}]{gotze.wolfle:1972}
\bibinfo{author}{\bibfnamefont{W.}~\bibnamefont{{G{\"o}tze}}} \bibnamefont{and}
  \bibinfo{author}{\bibfnamefont{P.}~\bibnamefont{{W{\"o}lfle}}},
  \bibinfo{journal}{\prb} \textbf{\bibinfo{volume}{6}}, \bibinfo{pages}{1226}
  (\bibinfo{year}{1972}).

\bibitem[{\citenamefont{Giamarchi}(1991)}]{giamarchi:1991}
\bibinfo{author}{\bibfnamefont{T.}~\bibnamefont{Giamarchi}},
  \bibinfo{journal}{\prb} \textbf{\bibinfo{volume}{44}},
  \bibinfo{pages}{2905} (\bibinfo{year}{1991}).

\bibitem[{\citenamefont{{Rosch} and {Andrei}}(2000)}]{rosch.andrei:2000}
\bibinfo{author}{\bibfnamefont{A.}~\bibnamefont{{Rosch}}} \bibnamefont{and}
  \bibinfo{author}{\bibfnamefont{N.}~\bibnamefont{{Andrei}}},
  \bibinfo{journal}{\prl} \textbf{\bibinfo{volume}{85}},
  \bibinfo{pages}{1092} (\bibinfo{year}{2000}).

\bibitem[{\citenamefont{{Belitz}}(1984)}]{belitz:1984}
\bibinfo{author}{\bibfnamefont{D.}~\bibnamefont{{Belitz}}},
  \bibinfo{journal}{J.~Phys.~C}
  \textbf{\bibinfo{volume}{17}}, \bibinfo{pages}{2735} (\bibinfo{year}{1984}).

\bibitem[{\citenamefont{Boulat et~al.}(2006)\citenamefont{Boulat, Mehta,
  Andrei, Shimshoni, and Rosch}}]{boulat-2006}
\bibinfo{author}{\bibfnamefont{E.}~\bibnamefont{Boulat}},
  \bibinfo{author}{\bibfnamefont{P.}~\bibnamefont{Mehta}},
  \bibinfo{author}{\bibfnamefont{N.}~\bibnamefont{Andrei}},
  \bibinfo{author}{\bibfnamefont{E.}~\bibnamefont{Shimshoni}},
  \bibnamefont{and} \bibinfo{author}{\bibfnamefont{A.}~\bibnamefont{Rosch}},
  \eprint{e-print arXiv:cond-mat/0607837}  (\bibinfo{year}{2006}).
  
\bibitem[{\citenamefont{Sologubenko et~al.}(2007)\citenamefont{Sologubenko,
  Berggold, Lorenz, Rosch, Shimshoni, Phillips, and
  Turnbull}}]{sologubenko:2007}
\bibinfo{author}{\bibfnamefont{A.~V.} \bibnamefont{Sologubenko}},
  \bibinfo{author}{\bibfnamefont{K.}~\bibnamefont{Berggold}},
  \bibinfo{author}{\bibfnamefont{T.}~\bibnamefont{Lorenz}},
  \bibinfo{author}{\bibfnamefont{A.}~\bibnamefont{Rosch}},
  \bibinfo{author}{\bibfnamefont{E.}~\bibnamefont{Shimshoni}},
  \bibinfo{author}{\bibfnamefont{M.~D.} \bibnamefont{Phillips}},
  \bibnamefont{and} \bibinfo{author}{\bibfnamefont{M.~M.}
  \bibnamefont{Turnbull}}, \bibinfo{journal}{\prl}
  \textbf{\bibinfo{volume}{98}}, \bibinfo{pages}{107201}
  (\bibinfo{year}{2007}).

\bibitem[{\citenamefont{{Kohn}}(1964)}]{kohn:1964}
\bibinfo{author}{\bibfnamefont{W.}~\bibnamefont{{Kohn}}},
  \bibinfo{journal}{Physical Review} \textbf{\bibinfo{volume}{133}},
  \bibinfo{pages}{171} (\bibinfo{year}{1964}).

\bibitem[{\citenamefont{{Kadanoff} and {Martin}}(1963)}]{kadanoff.martin:1963}
\bibinfo{author}{\bibfnamefont{L.~P.} \bibnamefont{{Kadanoff}}}
  \bibnamefont{and} \bibinfo{author}{\bibfnamefont{P.~C.}
  \bibnamefont{{Martin}}}, \bibinfo{journal}{Ann.~Phys.~(N.Y.)}
  \textbf{\bibinfo{volume}{24}}, \bibinfo{pages}{419} (\bibinfo{year}{1963}).

\end{thebibliography}

\end{document}